\documentclass[aps,prb,twocolumn,showpacs,float]{revtex4}

\usepackage{graphicx,amsmath}
\usepackage{amsmath}
\usepackage{amsfonts}
\usepackage{amssymb}
\usepackage{graphicx}
\usepackage[justification=raggedright,singlelinecheck=false]{caption}

\newcommand{\abs}[1]{\ensuremath{\left\vert#1\right\vert}}
\renewcommand{\arccos}[1]{\ensuremath{\operatorname{arccos}\left(#1\right)}}
\newcommand{\arccosh}[1]{\ensuremath{\operatorname{arccosh}\left(#1\right)}}
\newcommand{\arcsinh}[1]{\ensuremath{\operatorname{arcsinh}\left(#1\right)}}

\newcommand{\arctanh}[1]{\ensuremath{\operatorname{arctanh}\left(#1\right)}}
\newcommand{\sign}[1]{\ensuremath{\operatorname{sign}\left(#1\right)}}
\newcommand{\heaviside}[1]{\ensuremath{\theta\left(#1\right)}}
\renewcommand{\Re}[1]{\ensuremath{\operatorname{Re} \left\{#1\right\} } }
\renewcommand{\Im}[1]{\ensuremath{\operatorname{Im} \left\{#1\right\} } }
\newcommand{\vekop}[1]{\boldsymbol{\hat{#1}} }
\newcommand{\vek}[1]{\boldsymbol{#1} }
\renewcommand{\vec}[1]{\boldsymbol{#1} }

\begin{document}

\title{Dynamical current-current susceptibility of gapped graphene}
\author{Andreas Scholz and John Schliemann}
\affiliation{Institute for Theoretical Physics, 
University of Regensburg, D-93040 Regensburg, Germany}
\date{\today}

\begin{abstract}
We present analytical expressions for the current-current correlation 
function in graphene for arbitrary frequency, wave vector, doping, and 
band gap induced by a mass term. 
In the static limit we analyze the Landau (orbital) 
and Pauli magnetization, as well as the Lindhard correction which 
describes Friedel and Ruderman-Kittel-Kasuya-Yosida oscillations. In the nonrelativistic 
limit we compare our results with the situation of the usual
two-dimensional electron gas (2DEG). We find that the orbital magnetic 
susceptibility (OMS) in gapped graphene is smeared out on an energy scale
given by the inverse mass. 
The  nonrelativistic limit of the plasmon dispersion and the 
Lindhard function reproduces the results of the 2DEG. The same 
conclusion is true for the Pauli part of the susceptibility. The peculiar 
band structure of gapped graphene leads to pseudospin paramagnetism 
and thus to a special form of the OMS.
\end{abstract}

\maketitle

\section{Introduction}
\label{sect:Introduction}
Graphene, which was first isolated in 2004,\cite{Novoselov} is the name of a 
monolayer of carbon atoms that are arranged in a hexagonal lattice. 
It differs compared to most two-dimensional systems by its relativistic 
energy-momentum relation and its nontrivial spinor structure,\cite{Wallace}
originating from the two-atomic Wigner-Seitz cell, and has remarkable 
electronic properties.\cite{Netro_2009} As a consequence, various 
effects like the anomalous quantum Hall effect \cite{Novoselov2, Zhang} or 
the Klein tunneling \cite{Young} have been discovered. In 1956, 
the orbital magnetization of two-dimensional graphite had already been calculated, 
\cite{McClure} indicating a strong diamagnetism in the undoped hexagonal 
lattice which was confirmed by recent experiments.\cite{Sepioni_2010}

In the present work, we study the response of the system to an electromagnetic 
potential in terms of the current-current correlation function. 
Similar studies have already been performed recently
regarding the density-density response of massless\cite{Wunsch, Sarma}
and massful\cite{Pyat} Dirac fermions, the Hall conductivity in the presence of
spin-orbit interactions\cite{Dyrdal_2009,Ingenhoven_2010} including self-energy
and vertex corrections,\cite{Sinitsyn_2006} and current-current correlations in
the absence of a mass term.\cite{Polini, Stauber}
Here we will generalize those results to the case of massive 
quasiparticles by taking into account a mass term which breaks sublattice symmetry, leading to a
gap between the valence and the conduction band.
It can occur due to different mechanisms, including
intrinsic spin-orbit coupling\cite{Min,Kane} (with a gap of $24\mu eV$),\cite{Gmitra_2009}
graphene placed on a suitable substrate ($E_g= 0.26eV$),\cite{Zhou} or adsorption of molecules
(with a gap of several electron volts).\cite{Graphan}
The current correlator is related to the polarization function, which was 
discussed earlier.\cite{Pyat, Wunsch, Sarma} Its limiting behavior 
determines the orbital and the Pauli magnetization, the plasmon spectra, and
the screening of electric or magnetic impurities. Without the mass term, 
the Landau magnetization is infinite for intrinsic graphene (i.e. zero
chemical potential, $\mu=0$) and 
zero for extrinsic graphene ($\mu\neq0$),\cite{Polini,Stauber,Ando,McClure,Sharapov_2004}
while the Pauli part vanishes for the former and is finite for the latter case.
As gapped graphene is similar to the two-dimensional electron gas (2DEG), 
we investigate in the nonrelativistic limit, i.e., the limit of
single-particle energies just above the band-gap parameter, and compare our 
results to that of the 2DEG.\cite{Stern}

The particular features of the density-density and current-current
correlation functions in graphene compared to the standard 2DEG rely
on the coupling of the orbital degrees of freedom to the
sublattice of pseudospin. On the other hand, semiconductor systems 
involving coupling to other internal degrees of freedom such as the physical 
electron spin have also been analyzed recently with similar aspects.
As examples, we mention studies of the dielectric function of
semiconductor 2DEGs with various types of spin-orbit coupling terms,
\cite{Pletyukhov06,Badalyan09} two-dimensional semiconductor hole systems,
\cite{Cheng01} and $p$-doped bulk semiconductors.
\cite{Schliemann} Moreover, the
dielectric function of graphene taking into account the full honeycomb
lattice structure (but not a mass term) was analyzed recently in 
Ref.~\cite{Stauber10}. Analytical expressions for the polarizability of graphene
with finite width of Landau levels, temperature, and mass term can be found in Ref. \cite{Pyat2010}.

This paper is organized as follows. After introducing in Sec.~\ref{sect:Model} 
the model Hamiltonian and pertaining quantities, we present in 
Sec.~\ref{sect:Results} analytical expressions for the 
longitudinal and transversal current-current correlation function. In 
Sec.~\ref{sect:Static_limit}, we focus on the static limit and determine
the orbital and Pauli magnetization. Moreover, we include 
many-body effects via random-phase 
approximation (RPA). In Sec.~\ref{sect:NR_limit}, we study the effect of 
an increasing mass term on typical quantities like the magnetic 
susceptibility, Friedel oscillations, and the plasmon spectra, and we
compare the results to the 2DEG. We close with conclusions in 
Sec.~\ref{sect:Conclusions}. In Appendix \ref{appendix:calculation}, 
one can find details of the calculation of the transversal susceptibility, 
while Appendix \ref{appendix:remarks} comments on the relation between 
current and density response.

\section{The model}
\label{sect:Model}
The atoms in graphene are arranged in a honeycomb lattice, where each unit 
cell contains two carbon atoms. The effective Hamiltonian near the corners 
of the Brillouin zone K/K', including a mass gap as 
well as finite doping, is given by, using standard notation
\begin{align}
\hat H_0 &= \sum\limits_{\vek k} \hat\Psi_{\vec k}^\dagger 
\begin{pmatrix} \mu + mv_F^2 & \pi_x\mp i \pi_y \\ 
\pi_x\pm i \pi_y & \mu -mv_F^2 \end{pmatrix} \hat\Psi_{\vec k} \label{Ham}
\end{align}
where we have introduced an electromagnetic vector potential via
$\vek\pi = \hbar \vek k + \vek A$. 
The upper (lower) sign refers to the point K (K'). The field 
operator is defined by 
$\vekop\Psi_{\vec k} = \begin{pmatrix} \hat a_{\vec k} \\ 
\hat b_{\vec k} \end{pmatrix} $, where $\hat a_{\vec k}$ and $\hat b_{\vec k}$ 
are the destruction operators of the Bloch states in the two sublattices. 
Concentrating on the K point [upper sign in Eq. (\ref{Ham})],
the eigenvalues and eigenspinors at zero vector potential
$\vec A=0$ are given by
\begin{align*}
E_\pm(\vec k) = \mu \pm\sqrt{ \left(\hbar v_Fk\right)^2 
+ \left(mv_F^2\right)^2} , \\
\left\vert \chi_{\pm}(\vec k) \right\rangle = \frac 1{\sqrt 2}
\begin{pmatrix} \sqrt{1\pm \frac{mv_F}{\sqrt{\left(\hbar k\right)^2
+\left(mv_F\right)^2}}} \\ 
\pm \sqrt{1\mp \frac{mv_F}{\sqrt{\left(\hbar k\right)^2
+\left(mv_F\right)^2}}} \: \frac{k_x+ik_y}k \end{pmatrix} ,
\end{align*}
with $k=\sqrt{k_x^2+k_y^2}$. The current operator follows from
\begin{align}
\vekop j_{\vec q} = \frac{\delta \hat H_0}{\delta \vek A} 
= v_F \sum\limits_{\vec k,\alpha,\beta} 
\vekop\Psi^\dagger_{\vec k-\vec q,\alpha} \vekop\sigma_{\alpha\beta} 
\vekop\Psi_{\vec k,\beta} \; , \label{current_op}
\end{align}
where $\vekop\sigma$ are the Pauli matrices. Equation (\ref{current_op}) 
is, up to $v_F$, equal to the pseudospin operator. The electric current can 
be connected with the vector potential via the correlation function 
$\chi_{\vec j_\mu \vec j_\nu}$, defined by the Kubo product \cite{Guil}
\begin{align*}
\chi_{AB}(\omega) = -\frac i{\hbar \mathcal A} \int_0^{\infty} dt 
\left\langle\left[\hat A(t),\hat B(0) \right]\right\rangle_0 e^{i\omega t} 
e^{-0t} \; .
\end{align*}
Our system is rotationally invariant and the current is thus a linear 
combination of a purely longitudinal and a transversal 
part ($q=\abs{\vek q}$):
\begin{align*}
\chi_{\vec j_\mu \vec j_\nu}(\vec q,\omega) &
= \frac{\vek q_\mu \vek q_\nu}{q^2} \chi_{\vec j \vec j}^L(q,\omega) \\
&\hspace*{1cm} + \left(\delta_{\mu\nu} - \frac{\vek q_\mu \vek q_\nu}{q^2} 
\right)  \chi_{\vec j \vec j}^T(q,\omega) \; .
\end{align*}
For a noninteracting system, $\chi_{\vec j_\mu \vec j_\nu}(\vec q,\omega)$ 
is given by
\begin{align}
\chi_{\vec j_\mu \vec j_\nu}(\vec q,\omega) = &-\frac {gv_F^2}{\mathcal A} \sum\limits_{\lambda_1,\lambda_2,\vec k} \frac{f(E_{\lambda_1}(\vec k)) - f(E_{\lambda_2}(\vec k+\vec q))} {\hbar\omega + E_{\lambda_1}(\vec k) - E_{\lambda_2}(\vec k+\vec q) + i0} \notag \\
& \times \left\langle \chi_{\lambda_1}\left(\vec k\right) \right\vert \vekop\sigma_\nu \left\vert \chi_{\lambda_2}\left(\vec k+\vec q\right) \right\rangle \notag \\
& \times \left\langle \chi_{\lambda_2}\left(\vec k +\vec q\right) \right\vert \vekop\sigma_\mu \left\vert \chi_{\lambda_1}\left(\vec k\right) \right\rangle  , \label{Def_K_mu_nu}
\end{align}
where $f(E)$ is the Fermi function, and $g$ counts orbital and spin 
degeneracies ($g=4$ in graphene).

The orbital magnetic susceptibility is given by the static transversal 
part of $\chi_{\vec j\vec j}$:\cite{Guil}
\begin{align}
\tilde\chi_{orb} = \frac {e^2}{c^2} \lim\limits_{q\to0} \frac{\chi_{\vec j\vec j}^T(q,0)}{q^2} \label{OMS_limit} \; .
\end{align}
\indent Because of the continuity equation, $i\partial_t \hat\rho_{\vec q} = \vek q\cdot \vekop j_{\vec q}$, the response to a scalar potential, i.e., the polarization function, is included in the current-current susceptibility. In graphene, this leads to the following relation (see Appendix \ref{appendix:remarks}):\cite{Polini, Stauber}
\begin{align}
\omega^2 \: \chi_{\rho\rho}(q,\omega) = q^2 \chi_{\vec j \vec j}^L(q,\omega) - \frac 1{\hbar \mathcal A} \left\langle\left[\vec q\cdot \vekop{j}_{\vec q},\hat\rho_{-\vec q}\right]\right\rangle_0 . \label{rel_dds_ccs2}
\end{align}
The second term on the right-hand side was calculated in Ref. \cite{Sabio} and reads
\begin{align*}
\frac 1{\hbar \mathcal A} \left\langle\left[\vec q\cdot \vekop{j}_{\vec q},\hat\rho_{-\vec q}\right]\right\rangle = \frac{g q^2 D}{8\pi\hbar^2} ,
\end{align*}
where $D$ is a cutoff parameter, which is usually chosen to be of the order of the inverse lattice constant.\cite{Peres} Note that the commutator is independent of the mass.

For the following, it is essential to distinguish the cases $mv_F^2 > \mu$ 
and $\mu > mv_F^2$. In the first, \textit{intrinsic} case, the Fermi energy 
lies between the two bands, while in the second, \textit{extrinsic} case, 
the Fermi energy lies either in the conduction or in the valence band. 
From here, we will omit the spatial indices and use the notation 
$\chi_{jj}\equiv \chi_{\vec j_x \vec j_x}$.\\

\begin{figure*}[t]
\centering
\begin{minipage}[b]{8 cm}
\includegraphics[scale=0.35]{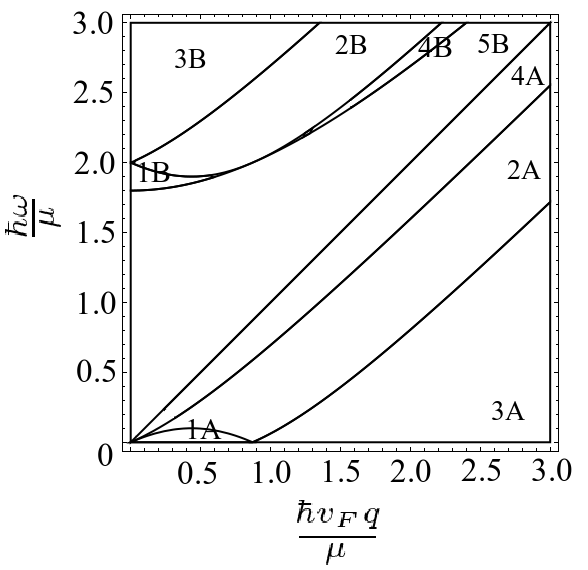}
\caption{The different regions related to the imaginary part of the 
current-current correlation function for the extrinsic case of
$mv_F^2/\mu = 0.9$. 
See Table \ref{Table_Regions} for the definitions of 1A-5B.}
\label{Regions}
\end{minipage}
\hspace*{1cm}
\begin{minipage}[b]{6 cm}
\captionof{table}{Definition of the different regions in the $q$-$\omega$-plane
related to the imaginary part of the current-current correlation function
in the extrinsic case, cf. Eq.(\ref{CCSuscept_Im_extr}).}
\begin{tabular}[t]{|l|l|}
\hline
1A: & $\hbar\omega < \mu - \sqrt{(\hbar v_F)^2(q-k_F)^2 + (mv_F^2)^2}$ \\ \hline 
1B: & $q<2k_F \wedge \; \sqrt{(\hbar v_Fq)^2+4(mv_F^2)^2} < \hbar\omega$ \\
    & $ \hspace*{0.8cm} < \mu + \sqrt{(\hbar v_F)^2 (q-k_F)^2 + (mv_F^2)^2}$ \\ \hline 
2A: & $\pm \mu \mp \sqrt{(\hbar v_F)^2 (q-k_F)^2 + (mv_F^2)^2} < \hbar\omega $\\
    & $\hspace*{0.8cm} < -\mu + \sqrt{(\hbar v_F)^2 (q+k_F)^2 + (mv_F^2)^2}$ \\ \hline 
2B: & $\mu + \sqrt{(\hbar v_F)^2 (q-k_F)^2 + (mv_F^2)^2} < \hbar\omega $ \\
    & $\hspace*{0.8cm} < \mu + \sqrt{(\hbar v_F)^2 (q+k_F)^2 + (mv_F^2)^2}$ \\ \hline 
3A: & $\hbar\omega < -\mu + \sqrt{(\hbar v_F)^2 (q-k_F)^2 + (mv_F^2)^2}$ \\ \hline 
3B: & $\hbar\omega > \mu + \sqrt{(\hbar v_F)^2(q+k_F)^2 + (mv_F^2)^2}$ \\ \hline 
4A: & $-\mu + \sqrt{(\hbar v_F)^2(q+k_F)^2 + (mv_F^2)^2} < \hbar\omega $ \\
    & $ \hspace*{0.8cm} < \hbar v_Fq$ \\ \hline 
4B: & $q>2k_F \; \wedge \; \sqrt{(\hbar v_Fq)^2 + 4(mv_F^2)^2} < \hbar\omega $ \\
    & $\hspace*{0.8cm} < \mu + \sqrt{(\hbar v_F)^2(q-k_F)^2 + (mv_F^2)^2}$ \\ \hline 
5B: & $\hbar v_Fq < \hbar\omega < \sqrt{(\hbar v_Fq)^2 + 4(mv_F^2)^2}$ \\
\hline
\end{tabular}
\label{Table_Regions}
\end{minipage}
\end{figure*}

\section{Results}
\label{sect:Results}
We restrict our discussions without loss of generality to positive 
frequencies $\omega$, chemical potentials $\mu$, and mass $m$. All other cases
follow from 
$\chi^{T/L}_{jj}(q,-\omega) = \left[\chi^{T/L}_{jj}(q,\omega)\right]^*$ and by 
observing that the results only depend on the absolute value of $\mu$ and $m^2$.

\subsection{Intrinsic case}
In the intrinsic case, only transitions from the valence into the conduction band contribute. As described in the last section, the longitudinal part, i.e., $\vek q = q \vekop x$, can be obtained from (\ref{rel_dds_ccs2}) and the density response given in Ref. \cite{Pyat}.
The longitudinal part was also directly calculated by the authors in order to check relation (\ref{rel_dds_ccs2}) for finite $m$. Because of the similarity to the transversal case, we restrict details of the calculation, given in the appendix, to the latter. The results are:
\begin{widetext}
\begin{align}
& \Im{\chi_{jj}^{L/T,int}\left(q,\omega\right)} = \frac{g\omega}{16\hbar} \sqrt{1-\left(\frac{v_Fq}\omega\right)^2}^{\,\mp1} \left(1+\frac{\left(2mv_F^2\right)^2}{\hbar^2\left(\omega^2-(v_Fq)^2\right)}\right) \heaviside{(\hbar\omega)^2-\left(\hbar v_Fq\right)^2-\left(2mv_F^2\right)^2} , \\[3ex]
& \Re{\chi_{jj}^{L,int}\left(q,\omega\right)} = \frac{g\left(D-2mv_F^2\right)}{8\pi\hbar^2} + \frac{gmv_F^2q^2}{4\pi\hbar^2\left(q^2-\omega^2/v_F^2\right)} + \frac{g\omega^2}{8\pi\hbar\sqrt{\abs{(v_Fq)^2-\omega^2}}} \left(1+\frac{\left(2mv_F^2\right)^2}{\hbar^2\left(\omega^2 - (v_Fq)^2\right)}\right) \notag \\
& \hspace*{1cm} \times \left[ \heaviside{v_Fq-\omega} \arccos{\frac{2mv_F^2}{\sqrt{\left(2mv_F^2\right)^2 + \hbar^2\left((v_Fq)^2-\omega^2\right)}}} - \heaviside{\omega-v_Fq} \arctanh{\frac {2mv_F^2}{\hbar\sqrt{\omega^2-(v_Fq)^2}}} \right] \label{CCSuscept_Im_intr} , \\[3ex]
&\Re{\chi_{jj}^{T,int}\left(q,\omega\right)} = \frac {g\left(D-2mv_F^2\right)}{8\pi\hbar^2} - \frac{g\sqrt{\abs{(v_Fq)^2-\omega^2}}} {8\pi\hbar} \left(1+\frac{\left(2mv_F^2\right)^2}{\hbar^2\left(\omega^2-(v_Fq)^2\right)} \right) \notag \\
& \hspace*{1cm} \times \left[ \heaviside{v_Fq-\omega} \arccos{\frac {2mv_F^2}{\sqrt{\left(2mv_F^2\right)^2+\hbar^2\left((v_Fq)^2-\omega^2\right)}} } -\heaviside{\omega-v_Fq} \arctanh{\frac {2mv_F^2}{\hbar\sqrt{\omega^2-(v_Fq)^2}} } \right] , \label{CCSuscept_Re_intr}
\end{align}
\end{widetext}
where $\heaviside{x}$ denotes the Heaviside step function.

\subsection{Extrinsic case}
We have two contributions for the extrinsic case. The first one is the 
undoped part where only interband transitions contribute (see above), 
while the second takes into account intraband transitions. Like in the 
intrinsic case, the longitudinal part is related to the density-density 
susceptibility via (\ref{rel_dds_ccs2}):
\begin{widetext}
\begin{align}
&\Im{\chi_{jj}^{L/T,ext}\left(q,\omega\right)} = \frac{g\omega}{16\pi\hbar} \sqrt{\abs{1-\left(\frac{v_Fq}{\omega}\right)^2}}^{\,\mp1}
\begin{cases}
G_>^{\mp}\left(\frac{2\mu+\hbar\omega}{\hbar v_Fq}\right) - G_>^{\mp}\left(\frac{2\mu-\hbar\omega}{\hbar v_Fq} \right) & \text{1A} \\
0 & \text{1B} \\
G_>^{\mp}\left(\frac{2\mu+\hbar\omega}{\hbar v_Fq} \right) & \text{2A} \\
\mp G_<^{\mp}\left(\frac{2\mu-\hbar\omega}{\hbar v_Fq} \right) & \text{2B} \\
0 & \text{3A} \\
\pi\left(1+\frac{\left(2mv_F^2\right)^2}{\hbar^2\left(\omega^2-(v_Fq)\right)^2} \right)  & \text{3B} \\
0 & \text{4A} \\
\pi\left(1+\frac{\left(2mv_F^2\right)^2}{\hbar^2\left(\omega^2-(v_Fq)\right)^2} \right)  & \text{4B} \\
0 & \text{5B ,} 
\end{cases} \label{CCSuscept_Im_extr} \\[2ex]
&\Re{\chi_{jj}^{L/T,ext}\left(q,\omega\right)} = \frac {gD}{8\pi\hbar^2} \pm \frac {g\mu\omega^2}{2\pi \left(\hbar v_Fq\right)^2} \mp \frac{g\omega}{16\pi\hbar} \sqrt{\abs{1 - \left(\frac{v_Fq}\omega\right)^2}}^{\,\mp1}  \, \begin{cases}
0 & \text{1A} \\
G_>^{\mp}\left(\frac{2\mu + \hbar\omega}{\hbar v_Fq}\right) - G_>^{\mp}\left(\frac{2\mu - \hbar\omega}{\hbar v_Fq} \right) & \text{1B}\\
\pm G_<^{\mp}\left( \frac{2\mu - \hbar\omega}{\hbar v_Fq} \right) & \text{2A}\\
G_>^{\mp}\left( \frac{2\mu + \hbar\omega}{\hbar v_Fq} \right) & \text{2B}\\
\pm G_<^{\mp}\left(\frac{2\mu - \hbar\omega}{\hbar v_Fq} \right) \pm G_<^{\mp}\left(\frac{2\mu + \hbar\omega}{\hbar v_Fq} \right) & \text{3A}\\
G_>^{\mp}\left( \frac{2\mu + \hbar\omega}{\hbar v_Fq} \right) - G_>^{\mp}\left(\frac{-2\mu + \hbar\omega}{\hbar v_Fq} \right) & \text{3B} \\
\pm G_<^{\mp}\left(\frac{2\mu - \hbar\omega}{\hbar v_Fq} \right) \mp G_<^{\mp}\left(\frac{2\mu + \hbar\omega}{\hbar v_Fq} \right) & \text{4A}\\ 
G_>^{\mp}\left( \frac{2\mu + \hbar\omega}{\hbar v_Fq} \right) + G_>^{\mp}\left(\frac{-2\mu + \hbar\omega}{\hbar v_Fq} \right) & \text{4B} \\
G_0^{\mp}\left(\frac{2\mu + \hbar\omega}{\hbar v_Fq} \right) - G_0^{\mp}\left(\frac{2\mu - \hbar\omega}{\hbar v_Fq} \right) & \text{5B .}
\end{cases} \label{CCSuscept_Re_extr}
\end{align}
\end{widetext}
Here we used the shorthand notation
\begin{align*}
&G_<^{\pm} = x\sqrt{x_0^2-x^2} \pm \left(2-x_0^2\right) \arccos{\frac x{x_0}} ,\\
&G_>^{\pm} = x\sqrt{x^2-x_0^2} \pm \left(2-x_0^2\right) \arccosh{\frac x{x_0}} ,\\
&G_0^{\pm} = x\sqrt{x^2-x_0^2} \pm \left(2-x_0^2\right) \arcsinh{\frac x{\abs{x_0}}} ,
\end{align*}
with 
$x_0=\sqrt{1+\frac{\left(2mv_F^2\right)^2}{\hbar^2\left((v_Fq)^2-\omega^2\right)}}$, and the regions (1A)-(5B), defined in Table \ref{Table_Regions} 
and Ref. \cite{Pyat}. The chemical potential is defined 
as $\mu=\sqrt{(\hbar v_Fk_F)^2+(mv_F^2)^2}$.
The above functions are one of the main results of this work. In the absence of a gap, we recover previous results.\cite{Polini, Stauber}
Figure \ref{Regions} illustrates the structure of the regions related to the 
imaginary part for the specific choice $mv_F^2/\mu = 0.9$.\\

\section{Static limit and magnetic susceptibility}
\label{sect:Static_limit}
\begin{figure*}
\leavevmode\includegraphics[scale=0.50]{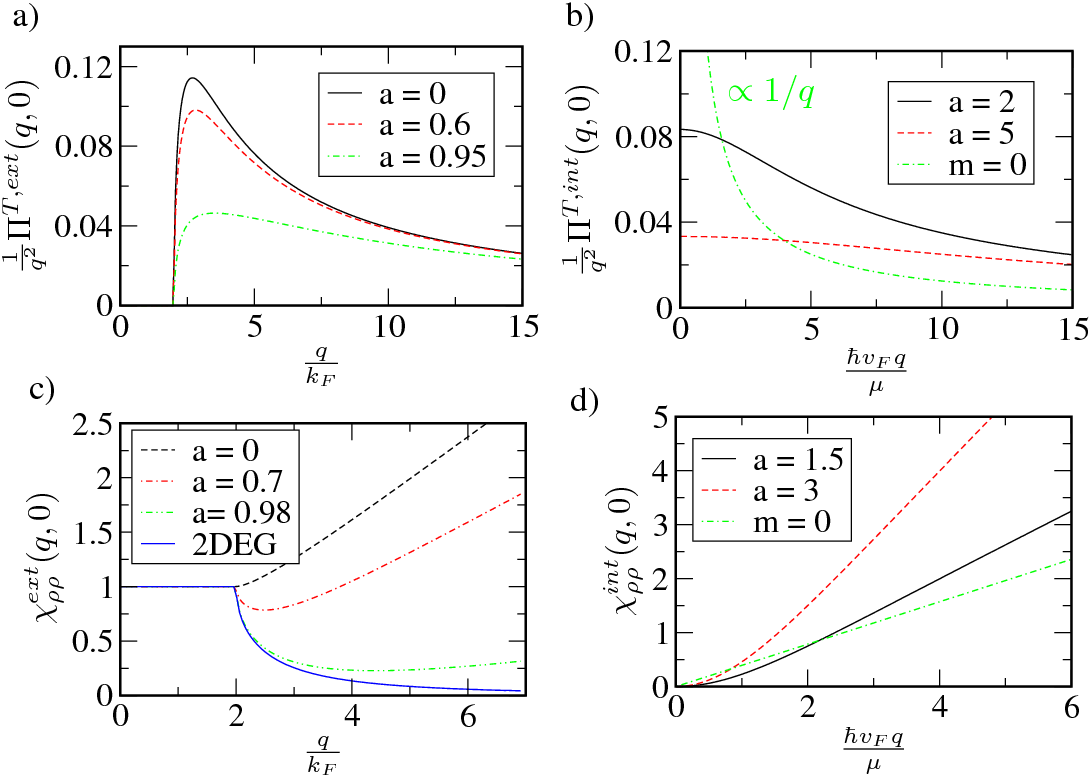}
\caption{(Color online) Static current correlation function for (\textbf{a)} extrinsic [Eq. (\ref{Static_extr})] and (\textbf{b)} intrinsic [Eq. (\ref{Static_intr})] graphene for different ratios $a\equiv mv_F^2 / \mu$ in units of $\frac{-gv_F}{2\pi\hbar k_F}$. Static polarization for (\textbf{c)} extrinsic [Eq. (\ref{polarization_static_ext})] and (\textbf{d)} intrinsic [Eq. (\ref{polarization_static_int})] graphene in units of $\frac {g\mu}{2\pi\hbar^2v_F^2}$.}
\label{Static_plot}
\end{figure*}

In the static limit, the purely real transversal susceptibility is given by
\begin{widetext}
\begin{align}
&\chi_{jj}^{T,int}\left(q,0\right) = \frac{gD}{8\pi\hbar^2} -\frac{gmv_F^2}{4\pi\hbar^2} - \frac{gv_Fq}{8\pi\hbar} \left(1-\left(\frac{2mv_F}{\hbar q}\right)^2 \right) \arccos{\frac{2mv_F}{\sqrt{(2mv_F)^2+(\hbar q)^2}}}  \label{Static_intr} , \\
&\chi_{jj}^{T,ext}\left(q,0\right) = \frac{gD}{8\pi\hbar^2} -\frac{gv_Fq}{8\pi\hbar} \left[\frac{2\mu}{\hbar v_Fq} \sqrt{1 - \left(\frac{2k_F}{q}\right)^2 } + \left(1 - \left(\frac{2mv_F}{\hbar q}\right)^2 \right) \cdot \arccos{\frac{2\mu}{\hbar v_Fqx_0}} \right] \heaviside{q-2k_F}\label{Static_extr} ,
\end{align}
\end{widetext}
while the longitudinal part vanishes, except for the constant term in front. 
Figure \ref{Static_plot} shows the function\\[2ex]
\begin{align*}
\frac 1{q^2} \Pi^T(q,0) = \frac 1{q^2} \left(\chi_{jj}^{T}\left(q,0\right) - \frac{gD}{8\pi\hbar^2} \right)
\end{align*}
for different values of $a\equiv mv_F^2/\mu$.

We now insert the above functions into (\ref{OMS_limit}). The intrinsic part,
\begin{align}
\tilde\chi_{orb}^{int} = -\frac {ge^2}{12\pi c^2m} \; , \label{OMS_gapped_int_graphene}
\end{align}
is finite and diamagnetic.
Compared to the gapless case, the orbital magnetic susceptibility (OMS) is smeared out on a scale of $1/m$. This broadening of $\tilde\chi_{orb}$ also occurs in the presence of disorder,\cite{Fukuyama}
as well as for finite temperature.\cite{McClure} From (\ref{Static_extr}), one can see that $\Pi^{T,ext}\left(q,0\right) = 0$ for $q<2k_F$ and thus
\begin{align*}
\tilde\chi_{orb}^{ext} = 0 \; ,
\end{align*}
which is the same as for ungapped graphene. The same result, namely,
\begin{align}
\tilde\chi_{orb} = -\frac {ge^2}{12\pi c^2m} \: \heaviside{mv_F^2 - \mu} \; ,
\end{align}
was obtained earlier by energy considerations.\cite{Sharapov_2004, Ando} The limit $m=0$ reproduces previous results:\cite{Polini, Stauber, McClure}
\begin{align*}
\tilde\chi_{orb} = -\frac{ge^2v_F^2}{6\pi c^2} \delta\left(\mu\right) \; .
\end{align*}
\indent The expressions for the magnetization given above are only valid for the noninteracting system. A simple way to include many-body effects is via the random-phase approximation.\cite{Guil} The OMS in RPA is given by
\begin{align*}
\tilde\chi_{orb}^{RPA} = \lim\limits_{q\to0} \frac{\frac 1{q^2} \Pi^{T}(q,0)}{1-\frac{2\pi e^2}{\epsilon_0 q}\Pi^{T}(q,0)} = -\frac {ge^2}{12\pi c^2m} \heaviside{mv_F^2 - \mu} ,
\end{align*}
where $\epsilon_0$ is the background dielectric constant. One can see that screening effects do not change the Landau part of the magnetization. Without a mass gap, the RPA result
\begin{align*}
\tilde\chi_{orb}^{RPA} = \left[1+\frac{g\pi e^2}{8\epsilon_0}\right]^{-1} \tilde\chi_{orb}
\end{align*}
yields to a renormalization, but the OMS remains infinite and zero, respectively. The situation changes, however, if one includes interaction effects in first-order perturbation theory beyond RPA, leading to paramagnetic behavior in doped graphene sheets.\cite{Polini2}\\
\indent The spin correlation function of a noninteracting system equals the density-density susceptibility.\cite{Guil} The Pauli contribution to the magnetization follows from the limit
\begin{align*}
\tilde\chi_P = \mu_B^2 \lim\limits_{q\to0} \chi_{S_zS_z}(q,0) \; ,
\end{align*}
where $\mu_B=\frac{e\hbar}{2m_0c}$ is the Bohr magneton and $m_0$ is the electrons bare mass. The static polarization reads\cite{Pyat}
\begin{widetext}
\begin{align}
&\chi^{int}_{\rho\rho}(q,0) = \frac{gm}{4\pi \hbar^2} - \frac{g\left(\left(\frac{2mv_F}{\hbar}\right)^2-q^2\right)}{8\pi \hbar v_Fq} \arccos{\frac{2mv_F}{\sqrt{(2mv_F)^2+(\hbar q)^2}}} \label{polarization_static_int} , \\[2ex]
&\chi^{ext}_{\rho\rho} \left(q,0 \right) = \frac{g\mu}{2\pi \hbar^2 v_F^2} \left\{ 1 - \frac 12 \left[ \sqrt{1-\left(\frac{2k_F}{q}\right)^2} - \frac{(\hbar v_Fq)^2-\left(2mv_F^2\right)^2}{2\hbar v_F\mu\, q} \arccos{\frac{2\mu}{\sqrt{\left(\hbar v_Fq\right)^2+\left(2mv_F^2\right)^2}}} \right] \, \heaviside{q-2k_F} \right\} . \label{polarization_static_ext}  
\end{align}
\end{widetext}
The Pauli part vanishes in the intrinsic case, reflecting the absence of states on the Fermi surface, while the extrinsic part is finite:
\begin{align}
\tilde\chi_P = \frac {ge^2\mu}{8\pi m_0^2 c^2 v_F^2} \: \heaviside{\mu-mv_F^2} \; . \label{Pauli_Suscep_ext_Graphene}
\end{align}
Figure \ref{Static_plot} displays the static polarization for different ratios $a=mv_F^2/\mu$. The limit $a\to1$ reflects the \textit{nonrelativistic case}.

\section{Nonrelativistic limit}
\label{sect:NR_limit}

\subsection{Magnetic susceptibility}
\begin{figure*}
\begin{center}
\leavevmode\includegraphics[scale=0.50]{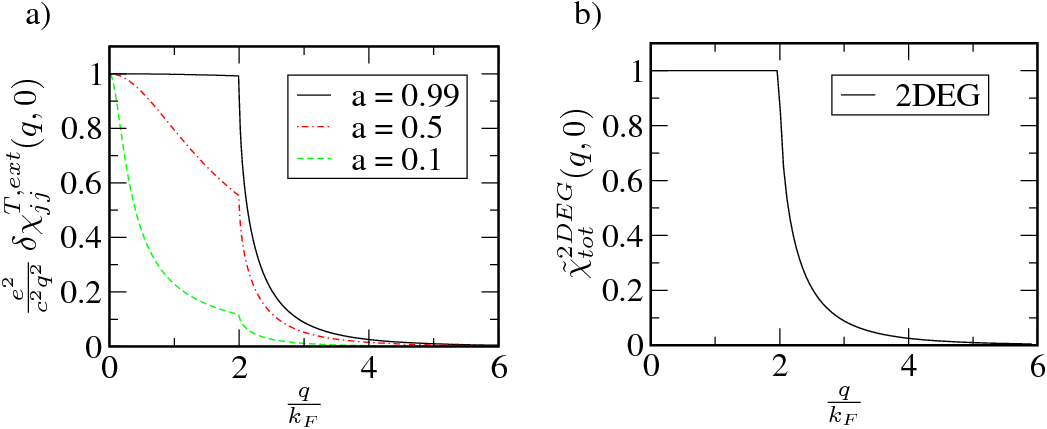}
\caption{(Color online) (\textbf{a)} Intraband part of the transversal current correlation function in graphene for different ratios $a\equiv mv_F^2/\mu$. (\textbf{b)} Sum of Pauli and Landau contribution in the 2DEG for the special case $m=m_0$. Both quantities are given in units of $\frac {ge^2}{12\pi c^2m}$.}
\label{2DEG_ges_Suscep}
\end{center}
\end{figure*}

The static transversal correlation function for the 2DEG,\cite{Guil}
\begin{align*}
\chi^{T,2DEG}_{jj}(q,0) = -\frac{gq^2}{24\pi m} \left[ 1 - \left(1-\frac{4k_F^2}{q^2}\right)^{3/2} \heaviside{q-2k_F} \right] ,
\end{align*}
leads to the OMS
\begin{align*}
\tilde\chi^{2DEG}_{orb} = -\frac{ge^2}{24\pi m c^2} \; ,
\end{align*}
where $g$ is a degeneracy factor. As described in the last section, the Pauli contribution to the total magnetization is given by the static polarization function \cite{Stern}
\begin{align*}
\chi_{\rho\rho}^{2DEG}(q,0) = \frac{gm}{2\pi \hbar^2} \left[1 - \heaviside{q-2k_F} \sqrt{1-\left(\frac{2k_F}q\right)^2} \; \right] ,
\end{align*}
and leads to
\begin{align*}
\tilde\chi_P^{2DEG} = \mu_B^2 \: \frac{gm}{2\pi \hbar^2} = \frac{ge^2m}{8\pi m_0^2 c^2} \; .
\end{align*}
Figure \ref{2DEG_ges_Suscep} displays the function 
\begin{align}
&\tilde\chi_{tot}^{2DEG}(q,0) = \mu_B^2 \,\chi_{\rho\rho}^{2DEG}(q,0) + \frac {e^2}{c^2q^2} \, \chi^{T,2DEG}_{jj}(q,0) \notag \\
&= \frac{ge^2}{12\pi mc^2} \left[1- \frac 32\heaviside{q-2k_F}\left\{\sqrt{1-\left(\frac {2k_F}q\right)^2} \right. \right. \label{Static_total_suscep_2DEG}\\
& \left. \hspace*{2cm} \left. - \frac 13\left(1-\left(\frac{2k_F}q\right)^2\right)^{3/2} \right\} \right] \notag
\end{align}
for the special case $m=m_0$. Its limit $q\to0$ determines the total magnetic susceptibility:
\begin{align}
\tilde\chi_{tot}^{2DEG} = \frac{ge^2m}{8\pi c^2m_0^2} \left(1-\frac13\left(\frac{m_0}m\right)^2\right) \stackrel{m=m_0} = \frac{ge^2}{12\pi c^2m} \; . \label{Suscept_2DEG_gesamt}
\end{align}

\indent Expanding the graphene Hamiltonian (\ref{Ham}) in the limit
$\pi/mv_F \ll1$, and eliminating the lower spinor component, one finds \cite{Ando}
\begin{align}
&\hat H_A = \frac{\vec\pi^2}{2m} + \frac \kappa2 g^* \mu_B B \label{graphene_Ham_NR} \; ,
\end{align}
where $g^*=2\frac {m_0}{m}$ is the effective Lande factor. $\kappa$ is dependent on the valley, i.e., $\kappa=-1$ for the K point and $\kappa=+1$ for K'.
Equation (\ref{graphene_Ham_NR}) is the well-known Hamiltonian of the 2DEG, including a Zeeman term which changes its sign by interchanging the two valleys. This Zeeman term, however, has nothing to do with the splitting of the energy levels due to the real spin, but is a truly band structure effect. Because of this, the second part of (\ref{graphene_Ham_NR}) is denoted as the \textit{pseudospin} Zeeman term.\cite{Ando} If we neglect states with negative energies, then the susceptibility associated with $\hat H_A$ is that of (\ref{Static_total_suscep_2DEG}), while the magnetization is given by (\ref{Suscept_2DEG_gesamt}). At the same time, the OMS of extrinsic graphene, i.e., for $\mu>mv_F^2$, including only intraband contributions, is given by the paramagnetic term
\begin{align}
\tilde\chi_{orb}^{intraband} = \frac {ge^2}{12\pi c^2m} \label{OMS_nur_intraband} \; ,
\end{align}
which means that the OMS of gapped graphene without hole states reproduces the total susceptibility of the 2DEG, i.e., the sum of the Pauli and the Landau part. Additionally, (\ref{Pauli_Suscep_ext_Graphene}) describes the Pauli part due to the \textit{real} spin. In the nonrelativistic limit $\mu\approx mv_F^2 + \frac {\hbar^2 k_F^2}{2m}$, Eq. (\ref{Pauli_Suscep_ext_Graphene}) reads
\begin{align}
\tilde\chi_P \approx \frac {ge^2m}{8\pi c^2m_0^2} \; , \label{Pauli_Suscep_ext_Graphene_NR_limit}
\end{align}
which is just the result of the 2DEG. Note that (\ref{Pauli_Suscep_ext_Graphene_NR_limit}) is true for extrinsic graphene with and without interband contributions.

\subsection{Friedel oscillations and plasmon dispersion}
Because of the divergent first derivative of the Lindhard correction (i.e., the static polarization) at $q=2k_F$ [see Fig. \ref{Static_plot}(c) for $a>0$], Friedel oscillations in gapped graphene behave differently compared to the gapless case, where the first derivative is finite but the second diverges [see Fig. \ref{Static_plot}(c) for $a=0$]. The system's reaction to charged impurities is described by \cite{Pyat}
\begin{align*}
\Phi_{total}(r) &= - \frac{Q\left(m{v_F}^2\right)^2}{\epsilon_0 a_0\mu^2} \cdot \frac {\left(2k_F\right)^2}{\left(2k_F+\frac 1{a_0}\right)^2} \cdot \frac {\sin{(2k_Fr)}}{\left(2k_Fr\right)^2} ,
\end{align*}
and is similar to the induced spin density $\delta m(r)$, which describes the interaction between magnetic moments, e.g., due to magnetic impurities:
\begin{align*}
\delta m\left(\vec r\,\right) &\propto -  \int \frac {d\,^2q}{\left(2\pi\right)^2} \cdot \chi_{\rho\rho}(q,0) \cdot e^{i\vec q \cdot \vec r} \notag \\
&= \frac {g\left(m{v_F}^2\right)^2 \cdot \left(2k_F\right)^2} {\left(2\pi v_F\right)^2 \hbar\mu} \cdot \frac {\sin{\left(2k_Fr\right)}}{\left(2k_Fr\right)^2} .
\end{align*}
Here, $a_0 = \frac {\epsilon_0\hbar^2{v_F}^2}{ge^2\mu}$ is an effective Bohr radius. In both cases, the nonrelativistic limit reproduces the result of the 2DEG,\cite{Stern}
\begin{align*}
\Phi_{total}(r) \approx -\frac{Q}{\epsilon_0 a_0} \cdot \frac {4k_F^2}{\left(2k_F+\frac 1{a_0}\right)^2} \cdot \frac {\sin{(2k_Fr)}}{\left(2k_Fr\right)^2}
\end{align*}
and\cite{Beal}
\begin{align*}
\delta m(r) \propto \frac {gm\left(2k_F\right)^2}{\left(2\pi\right)^2\hbar} \cdot \frac{\sin{\left(2k_Fr\right)}}{\left(2k_Fr\right)^2} ,
\end{align*}
while the massless case yields a different power law,\cite{Wunsch}
\begin{align*}
\Phi_{total}(r) / \delta m\left(\vec r\,\right) \; &\propto \frac{\sin{\left(2k_Fr\right)}}{(k_Fr)^3} .
\end{align*}
The long wavelength limit of the longitudinal susceptibility determines the dispersion of the collective modes.\cite{Guil} While plasmons are absent in intrinsic graphene, their dispersion for the extrinsic case reads \cite{Pyat}
\begin{align*}
&\hbar\omega_p \left(q\right) = \sqrt{\frac {ge^2 \mu}{2\epsilon_0} \cdot \left[1-\left(\frac{m{v_F}^2}{\mu}\right)^2 \right] \cdot q } .
\end{align*}
In the nonrelativistic limit, this can be approximated as ($n = \frac {g{k_F}^2}{4\pi}$)
\begin{align}
\hbar\omega_p(q) \approx \sqrt{\frac{2\pi \left(e\hbar \right)^2 n}{m\epsilon_0} \cdot q} 
\end{align}
which equals the 2DEG result \cite{Stern} and particularly shows the same $\sqrt{n}$ density dependence in contrast to the $n^{1/4}$ behavior of $m=0$.\cite{Sarma}

\subsection{Behavior near the threshold $\omega=v_Fq$}
The longitudinal current correlation function for graphene without bandgap is singular at $\omega=v_Fq$. This solely results from the linear dispersion relation. In gapped graphene, however, the singularity vanishes and the response quantities discussed in this work are smeared out on a scale of $1/m$. This is in accordance with the Lindhard function of the 2DEG,\cite{Stern} which is not singular at the threshold $\omega=v_Fq$. However, both the imaginary and the real part of $\chi_{\rho\rho}^{2DEG}$ are finite, while $\Im{\chi_{\rho\rho}(q,v_Fq)}$ vanishes in graphene [see Eq. (\ref{CCSuscept_Im_extr}) for 4A and 5B] and is thus in contrast to the 2DEG result. Furthermore, the real part at $\omega = v_Fq$ also differs from the result of the 2DEG.\\

\section{Conclusions}
\label{sect:Conclusions}
In this work, we have derived analytical expressions for the 
current-current correlation function of graphene for arbitrary frequencies, 
wave vectors, and doping, including a mass term whose sign depends on the 
sublattice. The static limit is of particular importance as it determines 
the magnetization of the system and the screening of impurities. The Landau 
magnetization of graphene without the mass term is proportional to the
$\delta$ function with respect to energy. 
As we have shown, this changes for finite masses in the 
intrinsic case, while 
the extrinsic result remains zero. The Pauli part of the susceptibility was 
found to be finite and positive for the extrinsic case and zero for the intrinsic 
case. As gapped graphene is formally quite similar to the 2DEG, we studied 
the nonrelativistic limit of the magnetization, the Friedel oscillations and 
the plasmon dispersion. We have demonstrated that all of these quantities, which 
follow directly from the transversal or longitudinal current correlation 
function, can reproduce the corresponding 2DEG results (e.g., the $n^{1/2}$ 
density dependence of the plasmon spectra or the $1/r^2$ decay law of the 
Friedel oscillations), but with one particularity, namely, the pseudospin 
Zeeman coupling.\\

\begin{acknowledgments}
We thank T. Stauber for useful discussions.
This work was supported by Deutsche Forschungsgemeinschaft via Grant No. GRK 1570.
\end{acknowledgments}

\appendix

\section{Details of the calculation of the transversal susceptibility}
\label{appendix:calculation}
In this section, we present details of the calculation of the transversal part of the current-current susceptibility. At zero temperature, Eq. (\ref{Def_K_mu_nu}) can be written as $\chi_{jj}(\vec q,\omega) = \xi_\mu^+ + \xi_\mu^- - \xi_D^- $ with
\begin{widetext}
\begin{align*}
\xi^\pm_\Lambda(\vec q,\omega) = -\frac {gv_F^2}{4\pi^2} & \int\limits d^2k \: \frac 12 \left(1 \mp \frac{(mv_F)^2-\hbar^2k^2\left(1-2\sin^2\varphi_{\vec k}\right) - \hbar^2qk\cos{\left(\varphi_{\vec k}+\varphi_{\vec q}\right)}} {E(k) \cdot E(\vec k+\vec q) } \right) \: \heaviside{\Lambda^2-(mv_F^2)^2-(\hbar v_Fk)^2} \\
&\times \left(\frac 1{\hbar\omega \mp E(\vec k+\vec q)+E(k)+i0} - \frac 1{\hbar\omega\pm E(\vec k+\vec q) - E(k) + i0} \right) .
\end{align*}
The plus (minus) sign corresponds to $\lambda_1=\lambda_2$ ($\lambda_1=-\lambda_2$). $\varphi_{\vec k}$ is the angle between $\vec k$ and the $\vekop x$ axis. For the longitudinal case ($\vec q=q\vekop x$), we obtain $\cos{\left(\varphi_{\vec k}+\varphi_{\vec q}\right)} = \cos{\varphi_{\vec k}}$, whereas the overlap for the transversal part ($\vec q=q\vekop y$) is given by $\cos{\left(\varphi_{\vec k}+\pi/2\right)} = -\sin{\varphi_{\vec k}}$. We now set for brevity $\hbar = v_F = 1$.

\subsection{Imaginary part}
We define the expression
\begin{align*}
I_{\sigma\tau}^\Lambda &= -\frac {g}{8\pi} \int\limits_0^\Lambda d^2k \left[1 - \sigma\frac{m^2-k^2\left(1-2 \sin^2\varphi \right) + qk\sin\varphi }{E(k) \cdot E(\vec k+\vec q)  } \right]  \delta\Big(\tau\omega - E(k) + \sigma E(\vec k+\vec q) \Big) \\
&= -\frac {g\sigma\sqrt{\omega^2-q^2}}{16\pi} \; \heaviside{1-\abs{\frac{\omega^2-q^2-2\tau\omega E(k)}{2qk}} } \:
\begin{cases}
G_<^+\left(\frac{2\sqrt{k^2+m^2} - \tau\omega}q\right) & \text{for } \omega > q ,\\
G_>^+\left(\frac{2\sqrt{k^2+m^2} - \tau\omega}q\right) & \text{for } q > \omega ,
\end{cases}
\end{align*}
where the functions $G_{>,<}^\pm$ and $x_0$ are defined in Sec. \ref{sect:Results}.\\
\indent The imaginary part for intrinsic graphene is given by
\begin{align*}
&\Im{\chi^{T,int}_{jj}\left(q,\omega\right)} = I_{--}^D - I_{-+}^D = \frac {g\sqrt{\omega^2-q^2}}{16} \times \\
& \hspace*{2cm} \times \left(1+\frac{4m^2}{\omega^2-q^2} \right) \heaviside{\omega^2-q^2-\left(2m\right)^2 } .
\end{align*}
\indent In the doped case, the upper integration limit is not a cutoff parameter but is the Fermi wave vector, and we thus need a distinction of cases as to whether $k$ is cut off by $k_F$ or not. For this, we define different regions,\cite{Pyat} which are given in Table \ref{Table_Regions} and displayed in Fig. \ref{Regions} in Sec. \ref{sect:Results}. The intraband contribution to the imaginary part reads
\begin{align*}
\Im{\delta \chi^{T,ext}_{jj}\left(q,\omega\right)} = \sum\limits_{\sigma,\tau=\pm1} \tau\, I_{\sigma\tau} = \frac{g\sqrt{\abs{\omega^2-q^2}}}{16\pi}
\begin{cases}
G_>^+\left(\frac{2\mu+\omega}q\right) - G_>^+\left(\frac{2\mu-\omega}q\right) & \text{1A} \\
-\pi\left(1+\frac{4m^2}{\omega^2-q^2} \right) & \text{1B} \\
G_>^+\left(\frac{2\mu+\omega}q\right) & \text{2A} \\
G_<^+\left(\frac{2\mu-\omega}q\right) -\pi\left(1+\frac{4m^2}{\omega^2-q^2} \right) & \text{2B} \\
0 & \text{3A} \\
0 & \text{3B} \\
0 & \text{4A} \\
0 & \text{4B} \\
0 & \text{5B} .
\end{cases}
\end{align*} 
The addition of the intrinsic part yields the final result given by Eq. (\ref{CCSuscept_Im_extr}).

\subsection{Real part}
The easiest way to find the real part of the intrinsic susceptibility is by using the Kramers-Kronig relation:
\begin{align*}
&\Re{\chi^{T,int}_{jj} \left(q,\omega\right) } = \frac 2\pi \mathcal P \int\limits_0^D dx \, \frac{x\Im{\chi^{T,int}_{jj}\left(q,x\right)} }{x^2-\omega^2} = \frac g{8\pi} \left(D-2m\right) - \frac{g\sqrt{\abs{q^2-\omega^2}}} {8\pi} \left(1+\frac{4m^2}{\omega^2-q^2} \right) \\
& \hspace*{4cm} \times\left[ \heaviside{q-\omega} \arccos{\frac {2m}{\sqrt{4m^2+q^2-\omega^2}} } - \heaviside{\omega-q} \arctanh{\frac {2m}{\sqrt{\omega^2-q^2}} } \right] .
\end{align*}
Note the cutoff-dependent part on the right-hand side. The real part of the extrinsic system is given as follows:
\begin{align*}
&\Re{\delta \chi^{T,ext}_{jj}\left(q,\omega\right)} = -\frac g{4\pi^2} \sum_{\tau=\pm1} \int d^2k \; \frac 12\left(1-\tau\frac{m^2-k^2\left(1-2\sin^2\varphi\right) + qk\sin\varphi} {E(k) \cdot E(\vec k+\vec q) } \right) \\
& \hspace*{6cm} \times \left[ \frac 1{\omega+E(\vec k) - \tau E(\vec k+\vec q) } - \frac 1{\omega - E(\vec k) + \tau E(\vec k+\vec q)} \right] \\
&= -\frac{g\omega^2 \left(\mu-m\right) }{2\pi q^2} - \frac{g\sqrt{\abs{\omega^2-q^2}}}{16\pi} \sum\limits_{\sigma=\pm1} \sign{ \frac{q^2-\omega^2}{2\omega}-\sigma E(k)} 
\begin{cases}
\left[ G_<^+\left(\frac{2E(k)+\sigma\omega}q\right) \right]_{k_1}^{k_2} & \text{for } q>\omega \\
\left[ G_>^+\left(\frac{2E(k)+\sigma\omega}q\right) \right]_{k_1}^{k_2}& \text{for } \omega^2>4m^2+q^2 \\
\left[ G_0^+\left(\frac{2E(k)+\sigma\omega}q\right) \right]_{k_1}^{k_2} & \text{for } q^2<\omega^2<4m^2+q^2 ,
\end{cases}
\end{align*}
where $k_1$ and $k_2$ are determined by the condition $\left(\frac{q^2-\omega^2-2\sigma\omega E(k)}{2qk}\right)^2 > 1$.\\
The intraband part of the susceptibility thus reads
\begin{align*}
\Re{\delta \chi^{T,ext}_{jj}\left(q,\omega\right)} &= -\frac{g\omega^2 \left(\mu-m\right) }{2\pi q^2} + \frac{g\sqrt{\abs{\omega^2-q^2}}}{16\pi} \begin{cases}
G_<^+\left(\frac{2m-\omega}q\right) + \sign{ \frac{q^2-\omega^2}{2\omega} - m}\cdot G_<^+\left(\frac{2m+\omega}q\right) & \text{A} \\
\sign{ \frac{q^2-\omega^2}{2\omega} + m}\cdot G_>^+\left(\frac{2m-\omega}q\right) - G_>^+\left(\frac{2m+\omega}q\right) & \text{1-4 B} \\
G_0^+\left(\frac{2m-\omega}q\right) - G_0^+\left(\frac{2m+\omega}q\right) & \text{5B}  
\end{cases} \\
+ &\frac{g\sqrt{\abs{\omega^2-q^2}}}{16\pi} 
\begin{cases}
0 & \text{1A} \\
G_>^+\left(\frac{2\mu+\omega}{q}\right) - G_>^+\left(\frac{2\mu-\omega}{q}\right)& \text{1B} \\
-G_<^+\left(\frac{2\mu-\omega}{q}\right) & \text{2A} \\
G_>^+\left(\frac{2\mu+\omega}{q}\right) & \text{2B} \\
-G_<^+\left(\frac{2\mu-\omega}{q}\right) - G_<^+\left(\frac{2\mu+\omega}{q}\right) &  \text{3A} \\
G_>^+\left(\frac{2\mu+\omega}{q}\right) - G_>^+\left(\frac{-2\mu+\omega}{q}\right)& \text{3B} \\
-G_<^+\left(\frac{2\mu-\omega}{q}\right) + G_<^+\left(\frac{2\mu+\omega}{q}\right) &  \text{4A} \\
G_>^+\left(\frac{-2\mu+\omega}{q}\right) + G_>^+\left(\frac{2\mu+\omega}{q}\right)& \text{4B} \\
G_0^+\left(\frac{2\mu+\omega}{q}\right) - G_0^+\left(\frac{2\mu-\omega}{q}\right) & \text{5B} \\
\end{cases}
\end{align*}
Adding the interband part from above yields the final result given by Eq. (\ref{CCSuscept_Re_extr}).

\section{Relation between current and density correlation function}
\label{appendix:remarks}
We define the four-current $\mathcal J^\mu = \begin{pmatrix} \hat \rho(\vec q,t) \\ - \vekop{j}(\vec q,t) \end{pmatrix}$. The four-current correlator can then be written as
\begin{align*}
&q_\mu \chi^{\mathcal J_\mu \mathcal J_\nu} (\vec q,\omega) \equiv - \frac{i}{\hbar \mathcal A} \int_{0}^\infty dt e^{i\omega t-0t} \left\{ \left\langle \left[\omega \hat\rho(\vec q,t), \mathcal J^\nu(-\vec q,0) \right] \right\rangle_0 -  \left\langle \left[\vec q\cdot \vekop j(\vec q,t), \mathcal J^\nu(-\vec q,0) \right] \right\rangle_0 \right\} = \\
&= - \frac{1}{\hbar \mathcal A} \left\{ \left[ e^{i\omega t - 0t} \left\langle \left[\hat\rho(\vec q,t), \mathcal J^\nu(-\vec q,0) \right] \right\rangle_0 \right]_{t=0}^\infty - \int_0^\infty dt e^{i\omega t - 0t} \left\langle \left[ \frac{\partial}{\partial t}\hat\rho(\vec q,t) + i \vec q\cdot \vekop j(\vec q,t), \left(1 + i\right) \mathcal J^\nu(-\vec q,0) \right] \right\rangle_0 \right\} \\ 
&= \frac{1}{\hbar \mathcal A} \left\langle \left[\hat\rho(\vec q,0), \mathcal J^\nu(-\vec q,0) \right] \right\rangle_0 ,
\end{align*}
where in the second line we used the continuity equation. This results in
\begin{align*}
&q^k \chi_{\vec j_k \vec j_l}(\vec q,\omega) q^l = \omega \chi_{\rho \vec j_k}(\vec q,\omega) q^l + \frac{1}{\hbar \mathcal A} \left\langle \left[\hat\rho_{\vec q}, \vec q \cdot \vekop j_{-\vec q} \right] \right\rangle_0 \\
&= \omega^2 \chi_{\rho \rho}(\vec q,\omega) -  \frac{1}{\hbar \mathcal A} \left\langle \left[\hat\rho_{\vec q}, \hat\rho_{-\vec q} \right] \right\rangle_0 + \frac{1}{\hbar \mathcal A} \left\langle \left[\hat\rho_{\vec q}, \vec q \cdot \vekop j_{-\vec q} \right] \right\rangle_0 .
\end{align*}
The second term on the right-hand side vanishes,\cite{Abedinpour_2011} while the third term needs special attention.\cite{Sabio} For a translational invariant system, i.e., $q^k \chi_{\vec j_k \vec j_l}(\vec q,\omega) q^l = q^2 \chi_{jj}^L$, one finally gets Eq. (\ref{rel_dds_ccs2}). In the \textit{n}DEG, the last term is exactly canceled by the diamagnetic contribution, i.e., $q_\mu \chi^{\mathcal J_\mu \mathcal J_\nu} = 0$, and thus $\left\langle \partial_\mu \mathcal J^\mu \right\rangle_0 = 0$. In our Dirac model, gauge invariance is broken because of the cutoff in the valence band. This can be seen, for example, by $\lim_{q\to0} \chi^L_{jj}(q,0) \neq 0$, which is unphysical, as a longitudinal static vector potential cannot induce a current. As stated in Ref. \cite{Stauber}, taking into account the full Brillouin zone leads to the cancellation of the commutator by a diamagnetic contribution (which is absent in the linearized model) and thus to $q_\mu \chi^{\mathcal J_\mu \mathcal J_\nu} q_\nu = 0$.
\end{widetext}

\end{document}